\documentstyle[a4,12pt]{article}
\def\y{\mbox{\bf 1}}
\def\Comp{\hbox{\bf C}}
\def\Real{\hbox{\bf R}}
\def\Inte{\hbox{\bf Z}}
\def\Natu{\hbox{\bf N}}

\def\Cs{$C^*$}
\catcode`\@=11
\def\defop#1{\expandafter
\def\csname#1\endcsname{\mathop{\rm#1}\nolimits}}
\defop{Map}  \defop{Mat}   \defop{Gr}   \defop{tr}
\defop{Tr}
\defop{Det}  \defop{str}   \defop{Str}  \defop{sdet}
\defop{Sdet}
\defop{Cliff}\defop{diag} \defop{sp}    \def\U{{\rm U}}
\defop{Cl}
\let\@@line\|       

\def\|#1>{\mathopen|#1\rangle}
\def\braket#1#2{\langle#1|#2\rangle}
\def\^#1#2#3{\langle#1|#2|#3\rangle}
\def\<#1>{\left\langle#1\right\rangle}
\def\=#1{{\bf#1}}
\def\-#1{{\bar{#1}}} 

\def\abs#1{\mathopen|#1\mathclose|}
\def\norm#1{\mathopen\@@line#1\mathclose\@@line}
\def\half{{1\over2}}
\def\slash{\@ifnextchar[{\@slash}{\@slash[\z@]}}
\def\@slash[#1]#2{
\setbox\z@\hbox{$#2$}\@tempdima\wd\z@\box\z@%
\@tempdimb#1 \advance\@tempdimb-\@tempdima \kern\@tempdimb
\hbox to\@tempdima{\hss\@makeslash\hss}}
\def\@makeslash{$/$}
\catcode`\@=12
\newcommand{\beq}{\begin{equation}}
\newcommand{\eeq}{\end{equation}}
\newcommand{\bea}{\begin{eqnarray}}
\newcommand{\eea}{\end{eqnarray}}
\newcommand{\ea}{\end{eqnarray*}}
\let\be\beq  \let\ee\eeq
\let\ba\bea  \let\ea\eea

\newcommand{\non}{\nonumber}
\let\oldref=\ref
\renewcommand{\ref}[1]{(\oldref{#1})}

\def\doo{\partial} \def\dii{\slash{\partial}}

\def\ot{\otimes} \def\op{\oplus}
\def\rar{\rightarrow}
\def\dd{{\rm d}} \def\eq{\mathop{\colon\!=}}

\def\matr#1#2#3#4{\mathop{{\left(
\begin{array}{cc}
 #1 & #2 \\ #3 & #4 \end{array} \right) }}}
\newcommand{\Df}[2]{\frac{{\rm d} #1}{{\rm d} #2}}
\newcommand{\df}[2]{\frac{\partial #1}{\partial #2}}
\newcommand{\dF}[2]{\frac{\delta #1}{\delta #2}}
\def\a{\alpha}   \def\b{\beta}
\def\g{\gamma}     \def\G{\Gamma}
\def\d{\delta}   \def\D{\Delta}
   \def\ve{\varepsilon}
    \def\E{\eta}
      \def\k{\kappa}
\def\l{\lambda}  \def\L{\Lambda}
\def\m{\mu}        \def\n{\nu}
\def\x{\xi}      
       
\def\s{\sigma}   \def\S{\Sigma}
       
\def\P{\Psi}     \def\th{\vartheta}
\def\th{\vartheta} 
     
\def\F{\Phi}       \def\c{\chi}
\def\o{\omega}   \def\O{\Omega}
   
\def\CA{{\cal A}} 
 \def\CD{{\cal D}}
\def\CE{{\cal E}} \def\CF{{\cal F}}
\def\CG{{\cal G}} \def\CH{{\cal H}}
 \def\CJ{{\cal J}}
 \def\CL{{\cal L}}
\def\CM{{\cal M}} 
\def\CO{{\cal O}} 
 
\def\CS{{\cal S}} 
 
\def\CW{{\cal W}} 
 
\let\Al\CA 
\let\Dir\CD  \let\M\CM
\def\mn{\mu\nu}
 
\def\As{\slash[1pt]{A}}

\begin{document}
\begin{titlepage}
\begin{flushright}
UUITP 17/94 \\
hep-th/9411075
\end{flushright}
\vskip 1.0 truecm
\begin{center}
{ \bf INDUCED QCD FROM THE NONCOMMUTATIVE GEOMETRY\\
OF A SUPERMANIFOLD }
\end{center}
\vskip 1.5cm
\begin{center}
{\bf Jussi Kalkkinen $^{*}$ } \\
\vskip 0.4cm
{\it Department of Theoretical Physics, Uppsala University
\\
P.O. Box 803,  S-75108 Uppsala, Sweden}
\vskip 0.4cm
\end{center}
\vskip 2.7cm

\rm
\noindent
We study the noncommutative geometry of a two-leaf
Parisi--Sourlas
supermanifold in Connes' formalism using different
$K$-cycles over the
Grassmann algebra valued functions on the supermanifold.
We find that
the curvature of the trivial noncommutative vector bundle
defines in
the simplest case the super Yang--Mills action coupled to
a scalar field.
By considering a
modified Dirac operator and a suitable limit of its
parameters we then
obtain an action that turns out to be the continuum
limit of the
induced QCD in Kazakov--Migdal model.
\vfill

\begin{flushleft}
\rule{5.1 in}{.007 in}\\
$^{*}$ { E-mail address:  jussi@rhea.teorfys.uu.se\\}
\end{flushleft}
\end{titlepage}

\section{Introduction}

Connes' theory of noncommutative geometry (NCG)
provides us with a
machinery to study the geometry of spaces, whose
structure may be
very different from that
of the ordinary smooth manifolds \cite{connes}.
It is believed that
NCG might give new ideas to the geometrical study
of quantum mechanics
and the quantum field theories.
Until now one has, however, worked mainly on
trying to explain some
features of classical field theories, as we
will also do in this paper.
As the result of these studies one now has
{\em e.g.~}a geometrical
interpretation for the Higgs field of the
Standard Model as a kind of
a connection field into a discrete direction:
the universe is supposed
in a sense to be an unconnected manifold.
An important result is also
the fact that the Standard Model can be expressed
in the terms of
NCG \cite{kastler}.  Some grand unified theories
\cite{frohlich1},
Chern--Simons theories \cite{frohlich3} and
Einstein--Hilbert gravity
coupled to a scalar field \cite{frohlich2}
have their equivalents as
the nontrivial geometry of some manifold in
the sense of NCG. Also
supersymmetric theories can be discussed in
this setting \cite{frohlich4},
but  problems in the definition of the
integral of differential
forms arise. These problems are present
also in our work, when we
study a supersymmetric manifold.

In this paper we first briefly review the
basic concepts of Connes' NCG,
and how classical field theory of scalar,
spinor and vector fields emerges
as a natural consequence of the geometry
and the symmetries of some given
manifold
\cite{frohlich3,varilly,jaffe}.
The emphasis is put on how to do the practical
calculations. We also
consider Feynman's path integral quantization
as a means to anticipate
the possible future implementation of quantum
field theories in NCG.
Next we take an example
of a noncommutative manifold, which can best
be described as a
two-leaf supermanifold (Parisi--Sourlas type).
We review briefly
the structure of the Clifford algebra on this
kind of a
manifold \cite{parisi,niemi}, and then apply
the machinery
of the NCG in order to study Yang--Mills
theories on the manifold.
The procedure here is similar to that
first presented in \cite{frohlich1}.
Connes' theory does leave some freedom to choose
the exact structure of
the Yang--Mills action, and some of the
possibilities to proceed are
examined. The result is in general a super
Yang--Mills theory coupled
to a scalar field. Using in a certain way
the freedom that we have
in the choice of the geometry, we consider
also a purely bosonic
theory that one obtains after integration
 over Grassmann coordinates.
In a particular limit of the parameters
we obtain the continuum limit
action of the Kazakov--Migdal model.

\section{Basic concepts}

The basic data that define a noncommutative
geometry \cite{connes,
varilly} are a \Cs-algebra $\CA$ and a
triple $(\CH,\CD,\G)$ called
a $K$-cycle. $\CH$ is a $\G$-graded Hilbert
space and the operator $\CD$,
often called the Dirac operator, is an odd
operator on $\CH$.
Elements of $\CA$ are suitable even operators
on $\CH$.

For example, the geometry of an ordinary,
{\em `commutative'},
manifold $\CM$
is obtained, if the algebra $\CA$ is the
algebra
$C^{\infty}(\CM, \Comp)$ of smooth, complex
valued functions of
the manifold, the Hilbert space $\CA$ is the
space of spinors on
this manifold, the grading is the Clifford
grading, and the Dirac
operator is the usual spin connection
$\dii + \slash{\o}$.
The studied manifold can be
{\em `noncommutative'}
in the sense that the algebra
$\CA$ is not commutative: An example
of this is the case where the algebra
above is tensored with the
space of complex $n \times n$-matrices
$\Mat(n, \Comp)$, and the
K-cycle is suitably modified. We will study
a special case of this this later in detail.

The basic tool for the study of geometry is
the differential algebra
$\O^*\CA$,
whose homogenous elements
\be
a^0 \dd a^1 \cdots \dd a^p \in \O^p \CA
\ee
are defined as elements of the tensor algebra
\be
\CA \ot \CA / \Comp \ot \cdots \ot \CA / \Comp.
\ee
Here two elements $a$ and $a'$ of the algebra
$ \CA $ represent the
same element in the quotient algebra $\CA / \Comp$ if
\be
a - a' = \l \y, ~~~~ \y \in \Al
\ee
is valid for some complex number $\l$ .

This differential algebra can be represented as operators on
$\CH$ using the homomorphism
\be
\pi(a^0 \dd a^1 \cdots \dd a^p) \eq a^0 i [\CD,a^1] \cdots
i [\CD,a^p].
\ee
In the case of a superalgebra, i.e.~when the
elements of $\CA$ are
not necessarily even, one should use a different
representation
\cite{jaffe}. Representation $\pi$
has the defect that the fact
$\pi(\a) = 0$ for some differential form
$\a$ does not imply $\pi(\dd \a)
= 0$.  Because we need a {\em differential}
representation of $\O^*\CA$ we
can thus not simply use  $\pi(\O^*\CA)$. A
suitable representation turns out
to be  the quotient algebra
\be
\O^*_{\CD}\CA \eq \pi(\O^*\Al)/ \pi ( \dd \ker \pi).
\ee
The representation mapping is denoted by
\be
\pi_{\CD} \colon \O^*\CA \rar \O^*_{\CD} \CA.
\ee

Connes defines the integration of differential
forms $\a \in \O^*\CA$
using the Dixmier
trace by the formula
\be
\int \a = \Tr_{\o} \pi_{\CD} (\a) \abs{\CD}^{-n},
\label{connint}
\ee
where $\o$ denotes a limiting procedure that picks
up the logarithmic
divergence of the conventional trace. The integer
$n$ is characteristic
to the chosen $K$-cycle and is associated with the
dimension of the
studied manifold by analog of the commutative case.
Once this integer
is found, one says that the $K$-cycle is
{\em $n$-summable}.
In the case of a superalgebra that  is not
$n$-summable for
any finite $n$ this definition can not be applied.
Instead, one can try
heat kernel regularization \cite{frohlich1,jaffe}
and define
\be
\int \a = \lim_{\b \rar 0+}
\Str \pi_{\CD}(\a)~e^{-\b \CD^2},
\ee
where $\Str$ is the usual  functional trace of
the Hilbert space.
In this article, the algebra $\CA$ is given using
a matrix algebra
of functions on a {\em a priori} know supermanifold.
We can thus
integrate differential forms using the formula
\be
\int \a \eq \int \dd^n x ~\dd^2 \E \sqrt{\mid
\sdet g \mid}
\Tr \pi_{\Dir}(\a(x,\E)),
\ee
where $\Tr$ is the ordinary, finite dimensional
matrix trace.
Other notation will be explained later.

The concept of a fiber bundle can also be introduced.
In the following,
only the trivial bundle $\CE \eq \Al$ is considered.
On this
generalization of a fiber bundle a covariant derivative
is given by
$\nabla = \dd + \a,~ \a \in \O^1\CA$.  The antihermitian
connection
one-form $\a$ defines the curvature two-form $\th \eq
\nabla^2 = \dd
\a + \a^2$.

\section{Field theory in NCG}

Once a fiber bundle structure has been
discovered, the natural thing to do is to start to study
the associated
Yang--Mills theory \cite{connes,varilly}
which, in general, is defined by the action
\be
\CS_{YM}[\a] \eq \int \th^2
\ee
 $\a$ being the connection 1-form.
This expression is not well defined as it stands,
because the
integration rules of the last section involve the
representation
$\pi_{\CD}$ and the way of choosing a representative
of an
element in the quotient algebra is not yet given.
This  ambiguity can be solved in two, equivalent ways
as it will turn out:
either directly fix a way
of  projecting any abstract element of the differential
algebra to its
unique representation operator \cite{frohlich3}, or define
the action
as the {\em minimum} of all possible choices of the
representative
\cite{varilly}.
The
first possibility amounts to defining an orthonormal
basis in the space of
differential forms with respect to the natural inner product
of $\O^*\CA$
\ba
\braket{\a}{\b} &=& \int \pi (\a \b^*).
\ea
Using the notation $\a = \a^{\perp} + \a^{\parallel} $
where $ \braket{\a^{\perp}}{ \b^{\parallel}} =0$ for all
$ \b^{\parallel}
\in d \ker \pi$ one then redefines
\be
\CS_{YM}[\a] = \int (\th^{\perp})^2.
\ee

The second possibility, which we will use here, amounts
to redefining
the action in the following way: Define first the
{\em `preaction'}
\be
\CS[\a] \eq \int \pi(\dd \a + \a^2)^2.
\ee
This is equivalent to modifying the integration rules
listed above such
that instead of $\pi_{\CD}$ there appears $\pi$. Now define
\be
\CS_{YM}[\a] \eq \inf_{\a' \in \O^1\CA} \Big\{
\CS[\a']~\Big|~\pi(\a-\a')
= 0 \Big\}.
\label{vaikutus}
\ee
Let us write the connection 1-form with the help of
elements of $\CA$
in form
\be
\a = \sum_{n,m} a^n \dd b^m,~~~~a^n,b^m \in \CA.
\ee
Both $\pi(\a)$ and $\pi(\dd \a + \a^2)$ are some
functionals of
the possibly very large set of fields $(a^n,b^m)$.
It is useful
to attempt to write down $\pi(\dd \a + \a^2)$ and
the preaction
with the help of  $\pi(\a)$ and  some functionally
independent
auxiliary field
$ P $. The Yang--Mills action is then the preaction
evaluated for
the fields $ ( \pi(\a), P^0  )$, where the fields
$ P^0 $ is the
solution of the functional equation
\be
\left. \dF{ S[\pi(\a), P]} {P} \right|_{P=P^0}=0.
\label{absel}
\ee
If the preaction turns out to be of the form
$\CS[\a] = \int \Tr
(\CL[\pi(\a)] + P)^2$
as it will in our later application, the
Euler--Lagrange equation
\ref{absel}
just implies that the effective lagrangian is
the projection of
$\CL[\pi(\a)]$
to the orthogonal direction with respect to
 the space of auxiliary
fields $P$.
This establishes the equivalence of the two methods.
Moreover, we notice that the Yang--Mills action
thus depends only
on rather $\pi(\a)$ than $\a$.

Fermions can be introduced to the theory as
vectors of $\CH$.
A suitable
action is
\be
\CS_{F}[\P,\a] = \^{\P}{\CD - i \pi(\a)}{\P},
\ee
where $\langle \cdot | \cdot\rangle $ is a
scalar product in $\CH$.

Both actions $\CS_{YM}$ and $\CS_{F}$ are
invariant under unitary
transformations of $\CE$. These
transformations turn out to be usual
local gauge transformations. Under such
a transformation
\be
u \in \U(\CE) = \U(\CA) = \{ v \in
\CA ~|~ v^*v = vv^* = \y \}
\ee
the transformation rules for the objects
introduced before
are
\ba
\g_u(\P) &=& u \P \\
\g_u(\a) &=& u \a u^* + u \dd u^* \\
\g_u(\th)&=& u \th u^*.
\ea
It will turn out to be useful to find the
transformation rule of the connection form
in terms of elements of
$\CA$. Let
\be
\a = \sum_{n,m} a^n \dd b^m,~~~~a^n,b^m \in
\CA.
\ee
Here the summation runs over a suitable set
of elements of $\CA$ that
does not necessarily need to be any kind of
a basis of the product
algebra $\CA
\ot \CA / \Comp$.  The index set can be large,
since for example on a
circle ${\rm S}^1$ we can use the decomposition
\be
 	a^n \dd b^m = e^{i n x } \ot e^{i
m y };~~~~n,m \in \Inte,
m \neq
0;~~ x,y \in [0,2\pi[.
\ee
In the following, only one index is used to
enumerate elements of $\O
\CA$ for convenience.  Adding to any $\a$ given
in this form the term
$(\L - \sum a^n \dd b^n) \dd \y$ for any $\L \in
\CA$ does not change
its value in
$\O^* \CA$ , but now the new coefficients
$a^n,b^n$ satisfy
\be
\sum_n a^n b^n = \L \label{lambda}.
\ee
In the case $\L = \y$ transformation properties
for the constituents of $\a$ reduce to
\ba
\g_u(a^n) &=& u a^n \\
\g_u(b^n) &=&  b^n u^*.
\ea

By now, there is no established way of
formulating  quantum field theories
in NCG. However,  since we do know how
to obtain classical field theories,
we can always quantize them in Feynman's
 path integral formalism. Because
the Yang--Mills and the fermion action only
depend on $\pi(\a)$ rather
than $\a$, as was previously shown,
one should attempt to define the generating
functional by
\be
[\dd \P] ~\exp \Big(-\CS_{YM}[\a]
\int  \pi(\a)^* \CJ_{F} \Big),
\CW = \int [\dd \pi(\a)] [\dd \P] ~\exp
\Big(-\CS_{YM}[\a]
-\CS_{F}[\a,\P] \Big).
\ee
are some elements of
When the connection one form is normalized
such that $\L = \y$ in formula
\ref{lambda} one can, as was shown, treat
the  $\U(\CE)$ invariance
of the original action as the local gauge
invariance of the spinors $\P$.
The gauge field is naturally $\pi(\a)$.
The way we choose to represent the
algebra $\CA$ in Hilbert space $\CH$ thus
fixes the gauge group.  One has
to remove this invariance of the integrand
by inserting a gauge condition
and a Faddeev--Popov determinant. However,
this naive quantization method
does not profit from any of the various
means the NCG would provide for
better understanding of quantum field
theories. The purpose of the
treatment above is merely to demonstrate
the significance of the
representation $\pi(\a)$ compared to the
abstract 1-form $\a$.

\section{The two-leaf supermanifold $\CM$}

The rest of the paper is devoted for studying
a special case of
noncommutative geometry. The starting data are
a supermanifold
$\CM$, its flat spin connection and complex
valued $n \times n$-matrix
functions defined on it. With these tools
we define the noncommutative manifold we
are about to study in the form
of a
$K$-cycle $(\CH,\CD,\G)$ over the algebra $\CA$.
We will also consider
the trivial fiber bundle $\CE = \CA$ on this
noncommutative manifold.

Let us start with the algebra of matrix-valued
functions on the given
supermanifold \be
\CA_0 = \Big[ \Mat(m,\Comp) \op \Mat(m',\Comp)
\Big] \ot C^{\infty}
(\CM,\CG),
\ee
where $m,m' \in \Natu$, $\CM$ is a Parisi--Sourlas
supermanifold with
local coordinates
\be (x^1,\cdots,x^n;\E^1,\E^2) \in \CM,
\ee
and $\CG$ is the (local) Grassmann algebra
generated by
$\E^1$ and $\E^2$. The Grassmann grading operator
that commutes with
bosonic numbers and anticommutes with fermionic
numbers is denoted
by $\g_g$.
This algebra can be geometrically interpreted as
the matrix-valued
function algebra
of a {\em two-leaf supermanifold}, if one reads
off the value of the
function on the first leaf from the left side of
the direct sum mark
and the value on the second leaf {}from the
right side.

Only a subalgebra $\CA \subset \CA_0$ of
this general case is
used here: First, the two leaves of the
manifold are associated with
each other in the sense that the value of
all allowed functions is the
same in corresponding points on the two
copies. As a  consequence of this
operation we get $m = m'$ and a complete
symmetry in indices that
distinguish between the two manifolds.
Secondly, let us restrict the
calculations to concern only the $\g_g$-even
elements.  The first
restriction can be physically understood in
such a way that it is
impossible to say on which leaf events take
place even if it is still
possible to
observe the existence of the gap between the
leaves. The second
restriction is imposed to obtain the right
grading for the naturally
arising connection fields.

The supermanifold $\CM$ is locally a direct
product of a bosonic manifold
with local coordinates $(x^1, \cdots, x^n)
\in \CM_b$ and a fermionic
manifold $(\E^1,\E^2) \in
\CM_f$ whose  metrics are $(g_{ij})$ and
$(g_{IJ})$
respectively\footnote{The lower case Latin
indices refer to
coordinates on $\CM_b$ and the upper case
indices to those on
$\CM_f$. Greek indices are used when no
distinction is made.
}  \cite{parisi,niemi}. The bosonic  part $\CM_b$
is supposed to have a spin$^c$-structure,
so that one can consider spinors
on it. The metric $g_{ij}$ has an euclidean
signature. The metric on
$\CM_f$
is
\be
(g_{IJ}) = \matr{0}{i\D}{-i\D}{0},~~~~ \D
\in \Real,
\ee
so that the bilinear form $X \cdot
X = X^{\m}X^{\n}g_{\mn}$ is real.

The elements on the tangent bundle $\dd
/ \dd t \in T_p\CM $ are of the
form
\be
\Df{}{t} = X^i \doo_i + X^I \doo_I,
\ee
where the anticommutation relation
$\{ \g_g, X^I\} = 0$ applies in
order to keep the tangent vectors
bosonic. Using the tangent
bundle one can define the super
Clifford algebra and the Clifford
bundle $\Cliff(T\CM,g)$ by imposing
the condition
\be
\{ X,Y \} = - 2 X \cdot Y \label{cliffalg}
\ee
in the tensor algebra of $T \CM \ni X,Y$. The
bosonic part yields
\be
\{ \={e}^i, \={e}^j  \} = -2 g^{ij},
\ee
where $\{ \={e}^{\m} \}$ is a local basis of the
tangent
space.  The bosonic part of the Clifford algebra
has an irreducible,
antihermitian representation. The representation
matrices are denoted
here $\g^i$.  The corresponding chirality operator
is denoted $\g_c$.
As the components of the tangent vectors are
$\g_g$-odd so must the
corresponding Clifford matrices be. Thus the
matrices $\={e}^I$
anticommute with the coordinates $\E^I$ and
the components $X^I$.  As a
consequence of \ref{cliffalg} they thus
satisfy the algebra
\be
[ \={e}^I, \={e}^J ] = -2 g^{IJ}.
\ee
The representation matrices are denoted by
\be
\G^I \eq \frac{1}{\sqrt{\D}} \g_c \ot \g_g \ot a^I,
\ee
where the antihermitian matrices $a^I$ satisfy the
algebra
\be
[a^1,a^2] = - 2 i ~ \y_{\CF}.
\ee
The letter $\CF$ stands for the infinite dimensional
representation
space (Fock space) of matrices $a^I$.
In general, the traces in this space are infinite and
must thus
be regularized, but it turns out that the only trace
we need to calculate
is that of the unity operator.  This we normalize to one:
\be
\tr \y_{\CF} = 1
\ee
In graded commutators we need the symbol $m_{\m}$
which is defined
\ba
m_{\m} \eq \left\{
\begin{array}{l}
0, \mbox{if~} \m = 1,\cdots, n \\
1, \mbox{if~} \m = \E^1, \E^2
\end{array}
\right.
{}.
\ea
Define further $(-)^{\m} \eq (-1)^{m_{\m}} $, $(-)^{\mn}
\eq
(-1)^{m_{\m}m_{\n}} $ and
\be
\{ A^{\m} , B^{\n} ] \eq A^{\m} B^{\n} +
(-)^{\mn} B^{\n} A^{\m} .
\ee
Using this notation the antihermitian representation
matrices
of the generators of the Clifford superalgebra
$\Cliff(T\CM,g)$
satisfy the relation
\be
\{ ~\G^{\m}, \G^{\n} ~] = - 2 ~ g^{\mn},
\ee
where $\G^i \eq \g^i \ot \y_{\CG} \ot \y_{\CF}$.
The nilpotent and
hermitian grading operator is $\g_c \ot \y_{\CG}
\ot \y_{\CF}$.

To show that the definitions given above are self
consistent,
we give an explicit realization of the Grassmann
algebra.  For example,
the matrices $\slash{n}^1$ and $\slash{n}^2$ of
the
Clifford algebra $\Cliff ( \Real^4, \diag(-1,-1,1,1) ) $
generate the
Grassmann algebra, that is
\be
\{ \slash{n}^I,\slash{n}^J \} = 0,
\ee
if we choose  $n^1 = (1,0,1,0)$, $n^2 =
(0,1,0,1)$.  This Clifford algebra has a real,
antihermitian
representation on the exterior algebra of $\Real^4$.
Integration and
derivation can
be represented as matrix operations.
The trace $\int \dd^2 \E $ is
normalized such that
\be
\int \dd^2 \E  ~ \E^I \E_I \eq \sqrt{
\mid \det (g_{IJ}) \mid }  ~.
\ee
Integrals over $\y$ and $\E^I$ disappear.
The grading operator $\g_g$
is here the chirality operator. The reason
for choosing the metric
$\diag(-1,-1,1,1)$ is the fact that this is
a simple way to make the
Grassmann algebra
close under adjunction.

Let us next define the $K$-cycle $(\CH,\CD,\G)$.
The Hilbert space
here is
\be
\CH \eq  S(T\CM_b) \ot \CG \ot \CF \ot \Comp^{2m + 2k},
\ee
where $k \in \Natu$ and $ S(T\CM_b) $ is the
spinor bundle on $\M_b$.
In this space the operator $ \dii = \G^{\m} \doo_{\m}$
is a hermitian
operator and it satisfies the (ungraded) Leibnitz rule,
when the inner
product of $\CH$ is
\be
\langle \P, \F \rangle \eq \int \dd^n x \dd^2 \E ~\sqrt{
\mid \sdet g
\mid } ~ \Tr (\P^* \F) .
\ee

The Dirac operator $\CD$ can be defined in several ways.
The simplest
definition is
\be
\CD = \matr{\dii \ot \y_{m} \ot \y_k}
{\G \ot M \ot K} {\G \ot M \ot K} {\dii \ot \y_{m}
\ot \y_k}
\label{superd},
\ee
where the grading is $\G =\s^3 \ot \g_c \ot \g_g$
and where $M$ and
$K$ are some given constant hermitian matrices.
In the Standard
Model \cite{kastler}
$M$ is a fermion mass matrix and $K$ is the
Kobayashi--Maskawa matrix
that mixes fermion families with each other.
Matrix $M$ operates in
the same space as the elements of $\CA$. In
the present case, $K$
operates in $\Comp^k$ and it does not have
common indices with any
other object. We would have been able to
use the whole spin connection
$\dii + \slash{\o}$, but since $\o$
disappears from all calculations,
we omit this unnecessary complication.
This definition produces,
as we shall presently
see, the usual Parisi--Sourlas Yang--Mills
theory.

Because the calculation is not restricted
by summability assumptions,
one can as well study the operator
\be
\CD \eq \matr{\dii \ot \y_{m} \ot \y_k}
{\g_c \ot M \ot K} {\g_c \ot M \ot K} {\dii
\ot \y_{m} \ot \y_k},
\label{perusd}
\ee
which gives rise to a  different classical
field theory. Furthermore,
one can scale the components of $\CD$
by any ($\g_g$-even) function, here denoted
$\c$ and $\x$. This means
considering operator
\be
\CD = \matr{\x ~ \dii \ot \y_{m} \ot \y_k}
{\c~ \g_c \ot M \ot K} {\c ~\g_c \ot M \ot K}
{ \x ~\dii \ot \y_{m}
\ot \y_k}.
\label{meidan}
\ee
We assume definition \ref{perusd} and
comment later on the other two
possibilities.

\section{The super Yang--Mills theory
of $\CM$}

In this section the action \ref{vaikutus}
is calculated for the NCG
defined in the previous section and expressed
as a functional of
physical fields.  The procedure relies on the
methods that Chamseddine
{\em et al.}~presented in \cite{frohlich1}.
Let
\ba
 \a &=& \sum (a^n \ot \y_2) \dd (b^n \ot \y_2)
\in \O^1\CA \\
a^n,b^n &\in & \Mat(m,\Comp) \ot C^{\infty}
(\CM,\CG),~~
\sum a^n b^n = \y .
\ea
In what follows, tensor products and unity
operators are omitted where
there is no danger of confusion.  We will
write the curvature
$\pi(\dd \a + \a^2)$ down with the help of
free fields some of
which turn out to be
nondynamical and which do not appear in
$\pi(\a)$. These auxiliary
fields are
eliminated with the help of appropriate
Euler--Lagrange equations so
that the action $S[\a]$ is in its minimum
for a given $\pi(\a)$.

First,  the dynamical fields  are
\be
\pi(\a) = \sum_n a^n ~ i [\CD, b^n] = i
\matr{\As}{\g_cK H}{\g_cK H}{\As},
\ee
where  the notation
\ba
A_{\m} &\eq& \sum a^n \doo_{\m} b^n~~~\mbox{and}
\label{mitta} \\
H &\eq& \sum a^n [M,b^n]
\ea
is used. The gauge transformations of these fields
are
\ba
\g_u (A_{\m}) = u A_{\m} u^* + u \doo_{\m} u^* \\
\g_u (H+M) = u (H +M) u^*,
\ea
where $u \in \U(\Mat(m,\Comp) \ot C^{\infty} (\CM,\CG)) $.
Thus the
field $A_{\m}$ transforms like a gauge field and $H+M$
like a scalar
field in the adjoint representation of the gauge group.

Due to the fact that the elements of $\CA$ are $\g_g$-even,
the
components of the fields $A^{\m}$ satisfy the grading
condition
\be
[ A^{\m}_{ab}, A^{\n}_{cd} \} = 0
\ee
and $A$ can be interpreted as the Parisi--Sourlas gauge
field. It
transforms in the transformations of the orthosymplectic
group
OSp$(n|2)$ as a vector so that $\tr (A^{\m} A_{\m}) $ is
invariant.

Because $\pi$ is a homomorphism, one obtains
\be
\pi(\a^2) = - \matr{ \As^2 + K^2 H^2 }{ ~[~ \As, H  \}
K \g_c }
{ ~[~ \As, H  \} K \g_c } { \As^2 + K^2 H^2 }
\ee
and
\be
\pi(\dd \a) = - \matr{ \dii \As + Z + K^2 \Big( \{ M,H \}  -
Y \Big) }
{\Big( ~[~\As, M ~\}~ - \slash{P} + \dii H \Big) K \g_c }
{\Big( ~[~\As, M ~\}~ - \slash{P} + \dii H \Big) K \g_c }
{ \dii \As + Z + K^2 \Big( \{ M,H \}  - Y \Big) },
\ee
where
\ba
Z&\eq& \sum a^n \doo^{\m} \doo_{\m} b^n \\
Y &\eq& \sum a^n [M^2,b^n ] \\
\slash{P} &\eq& 2 \sum a^n M \G^I \doo_I b^n.
\ea
The fields $Z$, $Y$ and $\slash{P}$ are thus the
naturally appearing
auxiliary fields. The curvature is now
\be
\pi(\th) = - \matr{\dii \As + \As^2 + Z + K^2
\Big( \F^2 - \P \Big) }
{\dii \F + [ \As, \F \} - \slash{P} }{\dii \F +
[ \As, \F \} - \slash{P} }
{\dii \As + \As^2 + Z + K^2 \Big( \F^2 - \P \Big) }  ,
\label{kaare}
\ee
where the new fields are $\F \eq H + M$ and $\P
\eq Y + M^2$.  If one
had used the definition \ref{superd} for the
Dirac operator $\CD$, one
would not have obtained the field $\slash{P}$
at all, and the graded
commutator in the formula \ref{kaare} above
would be an ordinary
commutator.

The trace $\Tr$ in formula \ref{vaikutus} can
be calculated in each
tensor product space separately
\be
\Tr = \tr_m \ot \tr_k \ot \tr_{\CF} \ot \tr_{\Cl}
\ot \tr_2
\ee
in obvious notation.  All traces are normalized so
that the trace of
the unit operator is one. Define
\be
\S^{\mn} \eq \frac{1}{4}[ \G^{\m}, \G^{\n} \}
\ee
so that $\G^{\m} \G^{\n} = - g^{\mn} + 2 \S^{\mn}$,
and
\be
F_{\mn} \eq \doo_{\m} A_{\n} - (-)^{\mn} \doo_{\n}
A_{\m} + [ A_{\m},
A_{\n} \}.
\ee
One can use the orthogonality properties of the
bosonic $\g$-matrices
\ba
\tr_{{\rm Cl}} \y^2 &=& 1 \\
\tr_{{\rm Cl}} \g^i\g_c \g^j\g_c &=&
g^{ij} \\
\tr_{{\rm Cl}} \s^{ij} \s^{nm} &=&
\frac{1}{4}(g^{im}g^{jn}-g^{in}g^{jm}).
\ea
Define also
\ba
\a^{IJ} &\eq& \frac{1}{4\D} \tr \{ a^I, a^J \} \\
\a^{IJKL} &\eq& \frac{1}{16\D^2} \Big(  \tr
( \{ a^I, a^J \}  \{ a^K, a^L \} )
- \tr \{ a^I, a^J \} \tr \{ a^K, a^L\} \Big).
\ea
Note that $\tr_{\CF}(a^Ia^J) \tr_m C_I C_J =
\D \tr_m C^I C_I$.
In above notation this yields
\ba
\CL &=& - \frac{1}{2} \tr F^{\mn} F_{\mn}  +
( \a^{IJKL} + \half g^{KI} g^{LJ}) F_{IJ} F_{KL} +
\non  \\
& & + \tr K^2 \tr (\nabla \F + [ \=A,\F])^2 +
(\tr K^4 -
\tr^2 K^2) \tr
(\F^2- \P)^2 + \non \\ & & + \tr \Big( - \a^{IJ}
F_{IJ} - (-)^{\m}
\doo^{\m} A_{\m} - (-)^{\m} A^{\m} A_{\m} + Z +
\tr K^2 (\F^2 - \P) ^2
\Big) + \non \\ & & - \tr \Big( \doo^I \F + \{A^I,
\F \} - P^I \Big)
\Big( \doo_I \F + \{A_I, \F \} - P_I \Big),
\label{tiheys}
\ea
where $\nabla^i := \doo^i$ and $\CL \eq \Tr
\pi(\th)^2$ .

The next step is to eliminate the auxiliary
fields using their
Euler--Lagrange equations. Obviously, trying
to eliminate the field
$\P$  would kill the potential of the scalar
field. The only way
to save the potential is to constrain the
matrices $M$ to be hermitian
matrices with the property
\be
M^2 = \half \m^2 \y + \b M
\ee
for some real $\m$ and $\b$, because this
yields $\P = (1/2) ~ \m^2 \y
+ \b \F$.  The two equations of motion to
be satisfied are
\ba
\doo_{\m}   \df{\CL}{\doo_{\m} Z_{ab} }   -
\df{ \CL }{ Z_{ab} } = 0 \\
\doo_{\m}   \df{\CL}{\doo_{\m} P^I_{ab} }  +
\df{ \CL }{ P^I_{ab} } = 0
\ea
for all $I$ and $(a,b)$.
The resulting constraints are
\ba
 0 & = & \doo^I \F + \{A^I, \F \} - P^I
\label{eka} \\ 0 & = & -
 \a^{IJ} F_{IJ} - (-)^{\m} \doo^{\m} A_{\m} -
(-)^{\m} A^{\m} A_{\m} +
 Z + \tr K^2 (\F^2 - \P) . \label{zeeta}
\ea
Substituting equations (\oldref{eka} --
\oldref{zeeta}) to the
lagrangian density \ref{tiheys} one obtains
\ba
\CS_{YM} &=&  \int \dd^n x
\dd^2 \E ~ \sqrt{\mid \sdet g \mid} ~ \bigg(  -
\half  \tr \Big(  F^{\mn}
F_{\mn}
\Big) + \non \\
& & + \tr(K^2) \tr \Big( (D^i \F )^* D_i \F \Big)
+ \Big( \tr K^4 -
\tr^2 K^2 \Big)
\tr V(\F) \bigg) , \label{tulos}
\ea
where the potential is
\be
V(\F) \eq \F^4 - 2 \b \F^3 + ( \b^2 - \m^2 ) \F^2
+ \b \m^2 \F
\ee
and  the covariant derivative is
\be
D^{\m}  \eq \doo^{\m}  + [A^{\m}, ~~ ] .
\ee
Formula \ref{tulos} resulted from the the fact that
the second term in
\ref{tiheys} disappears identically. This happens,
if we demand that
the elements of $\CA$ are invariant under the
transformations of the
orthosymplectic group OSp$(n|2)$ and no (other)
Grassmann parameters
appear in $\CA$. Then the elements are of the
form
\be
a^n ( x^{\m} ) = a^n_0 ( x^i  x_i ) +  a^n_2
( x^i  x_i )  \E^I  \E_I
\in  \CA
\ee
and a part of the components of the field tensor
disappear, namely
$F_{IJ} = 0$.  If one can choose such a
regularization that
the coefficients
\be
 \a^{IJKL} + \half g^{KI} g^{LJ}
\ee
vanish
the same result \ref{tulos} is obtained.

Action \ref{tulos} is clearly positive definite,
because $\tr K^4 \geq
\tr^2K^2$ holds.
proportional to the identity.
The fields $A_{\m}$ are antihermitian and the
field $\F$ is
hermitian. The action is obviously gauge invariant.
In this model the
scalar fields do not propagate to the fermionic direction.
On the
classical level it is impossible to choose the parameters
so that the
self-interaction would disappear and the scalar fields
stay massive.

\vspace{4mm}
If the Dirac operator had been defined  by formula
\ref{superd} instead of
the formula \ref{perusd} the resulting action
would have been
\ba
S_{YM} &=& \int \dd^n x \dd^2 \E ~\sqrt{ \mid
\sdet g \mid} ~\bigg( - \half
\tr\Big( F^{\mn} F_{\mn} \Big) + \non \\ & & +
\tr(K^2) \tr \Big(
(D^{\m}\F )^* D_{\m} \F \Big) + \Big( \tr K^4 -
\tr^2 K^2 \Big) \tr
V(\F) \bigg), \label{oiksqcd}
\ea
which is the conventional super Yang--Mills theory
coupled to a scalar field.

\section{The induced QCD}

Kazakov and Migdal have studied \cite{kazakov} the
induced QCD
starting from the lagrangian density
\be
\CL_{KM} =  \frac{N}{g^2} \tr \Big( (\nabla \Phi +
i g ~[\=A, \Phi] )^2 +
m^2 \Phi^2 + \l_0 \Phi^4 \Big). \label{km}
\ee
The idea is that after the path integration over
the scalar field in
the generating functional
\be
\CW[J^i] = \int [\dd \F] [\dd A] \exp \Big( -\int
\dd^n x ~\sqrt{\mid \det
{g_{ij}} \mid} ~ ( \CL_{KM} -J^i A_i ) \Big)
\ee
the resulting effective theory would be QCD.  In
this section the possibility of obtaining this
lagrangian $\CL_{KM}$
{}from the noncommutative geometry of some space
is studied.

In the previous section a Yang--Mills theory was
derived from a
suitable noncommutative geometry. The most general
deformation of the conventional Yang--Mills theory
is obtained by
deforming the Dirac operator as in formula
\ref{meidan}.
A similar procedure presented by Chamseddine {\em et
al.}~in
\cite{frohlich2} gives rise to a theory that involves
a scalar field
in Einstein--Hilbert gravitation theory.

Let us now assume the definition \ref{meidan} for
the Dirac operator.
All the calculations of the previous sections go
through so that one
only needs to scale derivatives in the resulting
expression by $\x$
and $K$-matrices by $\c$. The action thus obtained
is
\ba
\CS_{YM} &=&  \int \dd^n x \dd^2 \E ~ \sqrt{ \mid
\sdet g \mid} ~ \bigg(  -
\half  \x^4 \tr \Big(  F^{\mn} F_{\mn}  \Big) +
\non \\
& & + \x^2 \c^2 \tr(K^2) \tr \Big( (D^i \F )^* D_i
\F \Big) + \c^4
\Big( \tr K^4 - \tr^2 K^2 \Big) \tr V(\F) \bigg) ,
\label{tulos2}
\ea
Special forms for the functions  $\x$ and $\c$ are
chosen,
namely
\ba
\x  &\eq& \ve~ e^{\E^I\E_I} + \CO(\ve^2)\\
\c  &\eq& \ve~ e^{\E^I\E_I/\ve^4} + \CO(\ve^2).
\ea
This yields further
\ba
\x^4 &=& \CO(\ve^4) \\
\x^2\c^2 &=& 2 \E^I \E_I + \CO(\ve^4) \\
\c^4 &=& 4 \E^I \E_I + \CO(\ve^4) .
\ea
In the notation
\ba
\F(x^{\m} )  &=&  \F_0(x^i) +  \F_2(x^i) \E^I \E_I
\\
A^{\n} (x^{\m} )  &=&  A_0^{\n}(x^i) +  A_2^{\n}
(x^i) \E^I \E_I  \\
D^i_0 &=& \doo^i  +  [A_0^i ,  ~~]
\ea
and after Grassmann integration we obtain the action
\be
\CS_{YM} =  \int \dd^n x \sqrt{ \mid \det g_{ij}
\mid }~ \tr
\Big(  2 \tr (K^2) ~ (D^i_0  \F_0 )  (D_{0i}  \F_0 )
+ 4 ( \tr K^4 -
\tr^2K^2 )~ V(\F_0) \Big) . \label{loppu}
\ee
The result \ref{loppu} is of the same form as the
continuum limit of
the induced QCD \ref{km}, when one redefines the
fields $\F_0
\longrightarrow \F $ and $A^{i} \longrightarrow i
g A$. The parameters
are given by formulae
\ba
\frac{N}{g^2} &=& 2 \tr K^2 \\
\l_0 &=& 2 \frac{\tr K^4 } {\tr K^2 } - 2 \tr K^2
\\
m^2 &=& \l_0 \m^2 \label{massa}.
\ea
In order to maintain the fields $\F$ and $A_i$
purely bosonic in gauge
transformations, the local gauge symmetry is
reduced to the group
$\U(m,\Comp)$.  This limit of the super
Yang--Mills theory thus
correctly produces the purely bosonic
Kazakov--Migdal theory.

It is essential that one studies the equations
of motion for some
finite $\ve$, because otherwise the constraint
\ref{zeeta} would look
different {\em c.f.}~\ref{absel}.  Other choices
of functions $\x$ and
$\c$ are possible. The
outcome of calculations is essentially the same
also with the choice $
\x = \x_0 \g_c \G^I \E_I + \k $ to the zeroth order in
$\k$.

The model has some features that are absent from
the Kazakov--Migdal
model. The calculation is a way to show the
existence of a hidden
supersymmetry in the model. The interpretation
of the matrix $K$ as a
Kobayashi--Maskawa matrix gives a motivation to
include
fermion families with nontrivial mixing.  The
formula \ref{massa}
means that the scalar particle having mass implies
it moves in a
nontrivial potential, at least on the classical level.
In the
quantum theory this does not
imply a massive particle having necessarily nontrivial
self-interaction:  it has been shown in the case of the
Standard model that
some classically predicted restrictions on the parameter
values do not
hold in a quantized theory \cite{alvarez}. Formula
\ref{massa}
has also an other interesting feature: namely, the
parameter $1/\m$
roughly describes the distance between the two leaves
of the manifold.
Thus the mass of the scalar particle is inversely
proportional to this
length scale, so that the proportionality constant
is the square root
of the coupling constant. This gives rise to the
geometric picture that
when the two leaves approach each other the mass of
the scalar particle
tends to infinity (for a given $\l$), and when the
distance between the
leaves  increases without bounds, the scalar becomes
massless. The
strength of the interaction is connected with the
{\em untriviality}
of the fermion mixing, or the Kobayashi--Maskawa
matrix.

\vspace{4mm}
Considering other functions $\x$ and $\c$ would
naturally lead to
different results.  For instance, as a by-product
of the calculation
above it is apparent that replacing $\x$ by $\c$
would naturally
produce an ordinary, purely bosonic Yang--Mills
theory.

Let us next find out what would have been the outcome
of the analysis,
if one would have considered the Dirac operator \ref{superd}.
This
amounts to scaling the action \ref{oiksqcd} with $\x$ and
$\c$ in the
above described manner. It yields
\ba
\CS_{YM} &=&  \int \dd^n x \dd^2 \E ~ \sqrt{\mid \sdet
g \mid} ~
\bigg(  -  \half
\x^4  \tr \Big( F^{\mn} F_{\mn}  \Big) + \non \\
& & + \x^2 \c^2 \tr(K^2) \tr \Big( (D^{\m} \F )^*
D_{\m} \F \Big) +
\c^4 \Big( \tr K^4 - \tr^2 K^2 \Big) \tr V(\F)
\bigg) .
\ea
After Grassmann integration the resulting action
differs from the
action \ref{loppu} only by
\be
\d \CS[ A, \F_0,\F_2] = 2 \tr K^2 \int  \dd^n x
\sqrt{ \mid \det g _{ij}
\mid }  ~ \tr (\F_2 + [A, \F_0] )^2 \label{lisuke}
\ee
where the notation is $A^I \eq A \E^I$.  The only
relevant matrix
indices are those in $\Comp^m$. Since both $\F_i$
and $-i A$ are
hermitian, we can choose as the basis  of the
traceless, hermitian
generators $T^a$ of SU$(m)$ and the unity operator.
The coefficients
$t^{abc}$ are the structure constants of SU$(m)$.
Let us path integrate over the field $A$. This
can be done after a change of variables
\be
B^c  \eq \F_2^c - t^{abc} A^a \F_0^b .
\ee
The jacobian associated to the transformation is
\be
\Det \left| \dF{i A^a(y)}{B^b(x)} \right| =
\Det^{-1} | i
t^{abc} ~ \F^c_0 (x) ~\d (x-y)| = \Det^{-1} |
\F_0  |  =
\exp \Big( -  \int \tr \ln \F_0 \Big) .
\label{detti}
\ee
It does not depend on $B$ so the Gaussian integration
can be
performed. One can also easily integrate over $\F_2$.
The effective
contribution to the action is thus
\be
\d \CS_{{\rm eff}} [\F_0] = \int \tr \ln \F_0.
\ee
The logarithmic term in the potential introduces
new interactions, and
the theory is not renormalizable. Thus the only
way to produce a
consistent bosonic theory with these
tools starting with a superspace would seem to
be to consider the Dirac
operator \ref{perusd}.

\section{Conclusions}

The studied NCG describes a two--leaf
supermanifold. The two leaves
are identified, so that it is not
possible to detect on which leaf
physical events take place. On this manifold
a noncommutative fiber
bundle structure has been studied. The bundle
structure has been
written in terms of ordinary principal bundles
and with the help of the
 associated connection field $A^{\m}$.  Connes'
formalism introduces
also a scalar field $\F$ that
appeared to be a connection field into the
discrete direction. The
mass of the scalar fields give the order of
the magnitude of the
distance between the leaves. The strength of
the self interaction of
the scalars depends only on the fermion
structure through the
$K$-matrix.

There has remained some freedom to choose
the Dirac operator in the
K-cycle. This freedom has been investigated
and some of its
consequences in the quantum field theory have
been pointed out.  The
curvature of a two-leaf Parisi-Sourlas
supermanifold has been shown to
define a super Yang--Mills theory in a
similar way as it does in a
purely bosonic case. This Yang--Mills
theory had the property that it
could be continuously deformed into a
purely bosonic action for the
induced QCD in the Kazakov--Migdal model.
The set of field theories that
can be formulated in the formalism of Connes'
NCG has thus been
extended to include also concrete theories on
supermanifolds.  The
treatment is, however, classical. The
quantization of field theories
in NCG remains an open problem.

\vspace{1 cm}
\noindent {\bf Acknowledgements}

\noindent
I thank A.~Niemi for suggesting the problem and for
discussions.


\end{document}